\documentclass[prb,twocolumn,showpacs]{revtex4}
\usepackage{graphicx} 
\newcommand{\etal}{{ \it et al. }}

\usepackage{amsmath,amsthm,amssymb,mathrsfs}
\usepackage{bm}

\begin{document}

\title{Thermally assisted spin transfer torque switching in synthetic free layers}

\author{Tomohiro Taniguchi$^{1,2}$ and Hiroshi Imamura$^{1}$}
 \affiliation{
 ${}^{1}$ 
 Nanosystem Research Institute, National Institute of Advanced Industrial Science and Technology, 
 Tsukuba, Ibaraki 305-8568, Japan, \\ 
 ${}^{2}$ 
 Institute of Applied Physics, University of Tsukuba, Tsukuba, Ibaraki 305-8573, Japan
 }

 \date{\today} 
 \begin{abstract}
  {
  We studied the magnetization reversal rates of 
  thermally assisted spin transfer torque switching 
  in a synthetic free layer theoretically. 
  By solving the Fokker-Planck equation, 
  we obtained
  the analytical expression of the switching probability 
  for both the weak and the strong coupling limit. 
  We found that the thermal stability is 
  proportional to $\Delta_{0}(1-I/I_{\rm c})^{2}$, 
  not $\Delta_{0}(1-I/I_{\rm c})$ 
  as argued by Koch \etal [Phys. Rev. Lett. ${\bm 92}$, 088302 (2004)], 
  where $I$ and $I_{\rm c}$ are 
  the electric current and 
  the critical current of spin transfer torque switching 
  at absolute zero temperature, 
  respectively. 
  The difference in the exponent of $(1-I/I_{\rm c})$ 
  leads to a significant underestimation of the thermal stability $\Delta_{0}$. 
  We also found that 
  fast switching is achieved 
  by choosing the appropriate direction 
  of the applied field. 
  }
 \end{abstract}

 \maketitle



\section{Introduction}
\label{sec:Introduction}

Spin transfer torque switching of the magnetization in 
ferromagnetic nanostructures
has been extensively studied 
both theoretically \cite{slonczewski89,slonczewski96,berger96} 
and experimentally \cite{katine00,kiselev03,huai04,fuchs04} 
because of its potential application 
to spin-electronics devices 
such as magnetic random access memory. 
For device applications, 
a thermal stability $\Delta_{0}=MH_{\rm an}V/(2k_{\rm B}T)$ of
more than 40 is required 
to guarantee retention time 
of longer than ten years, 
where $M$, $H_{\rm an}$, $V$, $k_{\rm B}$ and $T$ are 
the magnetization, the anisotropy field, the volume 
of the free layer, 
the Boltzmann constant, 
and the temperature, respectively. 

Recently, Hayakawa \etal \cite{hayakawa08} showed that 
the anti-ferromagnetically coupled synthetic free layer, 
CoFeB(2.6nm)/Ru(0.8nm)/CoFeB(2.6nm), 
in a CoFeB(fixed layer)/MgO/CoFeB/Ru/CoFeB 
magnetic tunnel junction 
shows a large thermal stability ($\Delta_{0}>80$)
compared to a single free layer. 
On the other hand, Yakata \etal \cite{yakata09,yakata10}
showed that 
a ferromagnetically coupled 
CoFeB/Ru/CoFeB synthetic free layer 
shows a large thermal stability 
($\Delta_{0}=146 \pm 29$ for CoFeB(2nm)/Ru(1.5nm)/CoFeB(2nm) 
and $248 \pm 60$ for CoFeB(2nm)/Ru(1.5nm)/CoFeB(4nm)) 
compared to the single 
and the anti-ferromagnetically coupled synthetic free layer. 
These intriguing results spurred us to 
study a thermally assisted spin transfer torque switching 
in synthetic free layer. 
In contrast to the large number of experimental studies 
\cite{hayakawa08,yakata09,yakata10}, 
few theoretical studies have been reported. 
Although the analytical expression 
of the switching rate 
of the thermally assisted spin transfer torque switching 
for the single free layer \cite{koch04,li04,apalkov05}, 
$P=1-\exp[-f_{0}t\exp\{-\Delta_{0}(1-I/I_{\rm c})(1-H_{\rm appl}/H_{\rm an})^{2}\}]$, 
has been widely used to fit the experiments 
[see Eqs. (1)-(3) in Refs. \cite{yakata09,yakata10}], 
where $I_{\rm c}$ is the critical current 
of the spin transfer torque switching at absolute zero temperature, 
it is not clear whether 
this single layer formula has validity 
when applied to a synthetic free layer. 
Thus, it is important to derive 
an analytical expression of the switching rate 
of the thermally assisted spin transfer torque switching 
for the synthetic free layer.


 In this paper, 
 we studied the thermally assisted 
 spin transfer torque switching rate 
 for a synthetic free layer 
 by solving the Fokker-Planck equation. 
 The analytical expressions of 
 the switching rate were obtained 
 for weak and strong coupling limits 
 of the F${}_{1}$ and F${}_{2}$ layers. 
 One of the main findings was that 
 the dependence of 
 the thermal stability $\Delta$ 
 on the current $I$ is given by 
 $\Delta \propto \Delta_{0}(1-I/I_{\rm c})^{2}$, 
 not $(1-I/I_{\rm c})$, 
 as argued by the previous authors: \cite{koch04}
 We emphasize that even for the single free layer 
 $\Delta$ is proportional to $(1-I/I_{\rm c})^{2}$. 
 The difference in the exponent of the factor $(1-I/I_{\rm c})$ 
 leads to a significant underestimation 
 of the thermal stability $\Delta_{0}$. 
 We found that 
 in the presence of the applied field $H_{\rm appl}$, 
 the switching times of the anti-ferromagnetically 
 and the ferromagnetically coupled synthetic layers are different, 
 and that fast switching is achieved 
 by choosing an appropriate direction 
 of $H_{\rm appl}$.



This paper is organized as follows. 
In Sec. \ref{sec:Fokker-Planck equation for synthetic free layer}, 
we introduce the Fokker-Planck equation 
for the synthetic free layer 
and its steady state solution. 
We also introduce approximations 
to obtain the analytical expression 
of the switching probability. 
In Secs. \ref{sec:Weak coupling limit} 
and \ref{sec:Strong coupling limit}, 
we present the calculation 
of the switching probability 
in the limits of 
the weak and the strong coupling 
of the F${}_{1}$ and F${}_{2}$ layers. 
In Sec. \ref{sec:Relation to other works}, 
we compare 
our results with 
those of other works.  
Section \ref{sec:Conclusions} summarizes our findings.



\begin{figure}
  \centerline{\includegraphics[width=1.0\columnwidth]{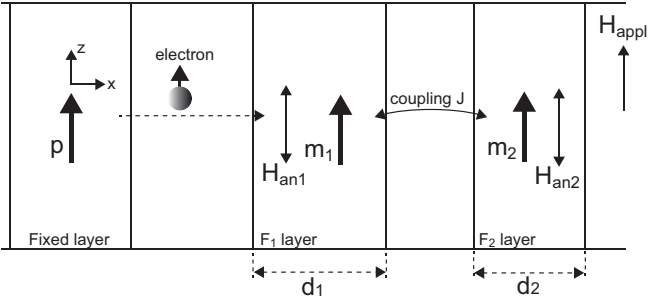}}
  \caption{
           The schematic view of the synthetic free layer 
           consisting of the F${}_{1}$ and F${}_{2}$ layers. 
           $\mathbf{m}_{k}$ and $\mathbf{p}$ are 
           the unit vectors along the directions of the magnetizations 
           in the F${}_{k}$ and the fixed layers, respectively, and 
           $d_{k}$ is the thickness of the F${}_{k}$ layer. 
           $H_{\rm appl}$, $H_{\rm an}$, and $J$ represent
           the applied field, the anisotropy field, 
           and the coupling between the F${}_{1}$ and F${}_{2}$ layers, respectively. 
           The flow of the electrons 
           along the $+x$ direction corresponds to 
           the negative electric current $I<0$. 
  }
  \label{fig:fig1}
\end{figure}



\section{Fokker-Planck equation for a synthetic free layer}
\label{sec:Fokker-Planck equation for synthetic free layer}

Let us first derive the Fokker-Planck equation 
for the synthetic free layer. 
The system we consider is schematically shown in Fig. \ref{fig:fig1}. 
The two ferromagnetic layers, F${}_{1}$ and F${}_{2}$, 
consist of a synthetic free layer 
with the coupling energy $-JS \mathbf{m}_{1}\cdot\mathbf{m}_{2}$. 
Here $\mathbf{m}_{k}=\mathbf{M}_{k}/M_{k}=(\sin\theta_{k}\cos\varphi_{k},\sin\theta_{k}\sin\varphi_{k},\cos\theta_{k})$ 
is the unit vector 
along the direction of the magnetization $\mathbf{M}_{k}$ 
of the F${}_{k}$ layer. 
$J$ and $S$ are the coupling energy per unit area 
and the cross-sectional area of the system, respectively. 
It should be noted that 
$J>0$ and $J<0$ correspond to 
the ferromagnetically coupled and antiferromagnetically coupled 
synthetic free layers, respectively. 
Although we consider the ferromagnetically coupled system below, 
our formalism is applicable to 
the antiferromagnetically coupled system 
by changing the sign of the coupling constant $J$. 
We assume the uniaxial anisotropy 
along the $z$ axis 
for both F${}_{1}$ and F${}_{2}$ layers, 
and the magnetizations 
$\mathbf{m}_{1}$ and $\mathbf{m}_{2}$ 
point to the positive $z$ direction 
in the initial states. 
We also assume that 
the external field $H_{\rm appl}$ is applied 
along the $z$ axis. 
Then, the total free energy $F$ of the F${}_{1}$ and F${}_{2}$ layers are given by 
\begin{equation}
\begin{split}
  F=&
  -M_{1}H_{\rm appl}V_{1}
  \cos\theta_{1}
  -
  \frac{1}{2}
  M_{1} H_{\rm an1}V_{1}
  \cos^{2}\theta_{1}
\\
  &-
  M_{2}H_{\rm appl}V_{2}
  \cos\theta_{2}
  -
  \frac{1}{2}
  M_{2} H_{\rm an2}V_{2}
  \cos^{2}\theta_{2}
\\
  &+
  2\pi M_{1}^{2}V_{1}
  (\sin\theta_{1}\cos\varphi_{1})^{2}
  +
  2\pi M_{2}^{2}V_{2}
  (\sin\theta_{2}\cos\varphi_{2})^{2}
\\
  &-JS 
  \left[
    \sin\theta_{1}
    \sin\theta_{2}
    \cos(\varphi_{1}-\varphi_{2})
    +
    \cos\theta_{1}
    \cos\theta_{2}
  \right],
  \label{eq:free_energy}
\end{split}
\end{equation}
where $H_{{\rm an}k}$, 
$V_{k}=Sd_{k}$ and $d_{k}$ are 
the uni-axial anisotropy field, 
the volume and the thickness of the F${}_{k}$ layer, 
respectively. 
The fifth and sixth terms 
in Eq. (\ref{eq:free_energy}) represent 
the magnetic energy due to the demagnetization field. 
We assume that $|H_{\rm appl}|<H_{{\rm an}k}$ 
to guarantee at least two local minima of the free energy. 
When $H_{Jk}\ll H_{{\rm an}k}$, 
the states
$(\mathbf{m}_{1},\mathbf{m}_{2})=(\mathbf{e}_{z},\mathbf{e}_{z})$,
$(\mathbf{e}_{z},-\mathbf{e}_{z})$,
$(-\mathbf{e}_{z},\mathbf{e}_{z})$, 
and $(-\mathbf{e}_{z},-\mathbf{e}_{z})$
correspond to the energy minima, 
where $H_{Jk}=J/(M_{k}d_{k})$. 
On the other hand, 
when $H_{Jk}\gg H_{{\rm an}k}$, 
the states
$(\mathbf{m}_{1},\mathbf{m}_{2})=(\mathbf{e}_{z},\mathbf{e}_{z})$ 
and $(-\mathbf{e}_{z},-\mathbf{e}_{z})$ 
correspond to the energy minima. 

The purpose of this paper is 
to investigate the switching rate 
of the magnetizations $\mathbf{m}_{1}$ and $\mathbf{m}_{2}$ 
from $\mathbf{m}_{1},\mathbf{m}_{2}=+\mathbf{e}_{z}$ 
to $\mathbf{m}_{1},\mathbf{m}_{2}=-\mathbf{e}_{z}$. 
Following Brown \cite{brown63}, 
we use the Fokker-Planck equation approach 
to calculate the switching probability per unit time, 
where the Fokker-Planck equation is derived 
from the equations of the motion of the magnetizations. 

We assume that 
the dynamics of the magnetizations of 
the F${}_{1}$ and F${}_{2}$ layers are described 
by the Landau-Lifshitz-Gilbert (LLG) equations. 
In general, 
the spin transfer torque acting on $\mathbf{m}_{1}$ 
arises from the spin currents 
injected from the fixed layer and the F${}_{2}$ layer. 
However, in a conventional synthetic free layer, 
the spacer layer between the F${}_{1}$ and F${}_{2}$ layers 
consists of Ru, 
whose spin diffusion length 
is comparable to its thickness \cite{comment1}; 
thus, the spin current 
injected from the F${}_{2}$ layer 
is negligible \cite{comment2}. 
Then, the LLG equation
of $\mathbf{m}_{1}$ is given by 
\begin{equation}
\begin{split}
  \frac{{\rm d}\mathbf{m}_{1}}{{\rm d}t}
  =&
  -\gamma_{1}
  \mathbf{m}_{1}
  \times
  \mathbf{H}_{1}
  +
  \gamma_{1}
  a_{J}
  \mathbf{m}_{1}
  \times
  (\mathbf{p}\times\mathbf{m}_{1})
\\
  &-
  \gamma_{1}
  \mathbf{m}_{1}
  \times
  \mathbf{h}_{1}
  +
  \alpha_{1}
  \mathbf{m}_{1}
  \times
  \frac{{\rm d}\mathbf{m}_{1}}{{\rm d}t}.
  \label{eq:LLG_F1}
\end{split}
\end{equation}
Similarly, 
the spin current injected from the F${}_{1}$ layer 
into the F${}_{2}$ layer 
is also negligible, 
and the LLG equation of $\mathbf{m}_{2}$ is given by 
\begin{equation}
\begin{split}
  \frac{{\rm d}\mathbf{m}_{2}}{{\rm d}t}
  =&
  -\gamma_{2}
  \mathbf{m}_{2}
  \times
  \mathbf{H}_{2}
  -
  \gamma_{2}
  \mathbf{m}_{2}
  \times
  \mathbf{h}_{2}
  +
  \alpha_{2}
  \mathbf{m}_{2}
  \times
  \frac{{\rm d}\mathbf{m}_{2}}{{\rm d}t},
  \label{eq:LLG_F2}
\end{split}
\end{equation}
where $\gamma_{k}$ and $\alpha_{k}$ 
are the gyromagnetic ratio and 
the Gilbert damping constant 
of the F${}_{k}$ layer, respectively. 
The magnetic field $\mathbf{H}_{k}$ 
acting on the magnetization $\mathbf{m}_{k}$ 
is defined by 
$\mathbf{H}_{k}=-(M_{k}V_{k})^{-1}\partial F/\partial\mathbf{m}_{k}$. 
$\mathbf{h}_{k}$ represents 
the random field 
on the F${}_{k}$ layer 
whose Cartesian components $h_{ki}$ ($i=x,y,z$) 
satisfy
\begin{equation}
  \langle h_{ki}(t) h_{k^{\prime}j}(t^{\prime}) \rangle 
  =
  \frac{2k_{\rm B}T\alpha_{k}}{\gamma_{k}M_{k}V_{k}}
  \delta_{kk^{\prime}}
  \delta_{ij}
  \delta(t-t^{\prime}),
\end{equation}
where $\langle \cdots \rangle$ means 
the ensemble average. 
Here we assume no correlation 
between the random fields 
acting on the F${}_{1}$ and F${}_{2}$ layers. 
The $a_{J}=\hbar \eta I/(2eM_{1}V_{1})$ term in Eq. (\ref{eq:LLG_F1}) represents 
the spin transfer torque 
due to the injection of the spin current 
from the fixed layer. 
Here $I$ is the electric current 
flowing along the $x$ axis. 
The positive electric current corresponds to the electron flow
along the $-x$ direction. 
$\eta$ is the spin polarization of the electric current 
which characterizes the strength of 
the spin transfer torque. 
The explicit form of $\eta$ depends on 
the theoretical model 
\cite{slonczewski96,brataas01,zhang02}, 
and, in general, 
depends on $\theta_{1}$. 
However, for simplicity, 
we assume that $\eta$ is constant 
(the dependence of $\eta$ on $\theta_{1}$ 
can be taken into account 
by replacing $a_{J}\cos\theta_{1}$ in Eq. (\ref{eq:effective_free_energy}) 
with $\int {\rm d} \cos\theta_{1} a_{J}$). 
$\mathbf{p}$ is 
the unit vector 
along the direction of the magnetization 
of the fixed layer. 

From the LLG equations 
(\ref{eq:LLG_F1}) and (\ref{eq:LLG_F2}), 
we obtain the Fokker-Planck equation 
for the probability distribution 
of the directions of the magnetizations, 
$W(\mathbf{m}_{1},\mathbf{m}_{2})$, 
which is given by \cite{brown63}
\begin{equation}
\begin{split}
  \frac{\partial W}{\partial t}
  &=
  \frac{\gamma_{1}}{M_{1}V_{1}}
  \frac{1}{\sin\theta_{1}}
  \frac{\partial}{\partial\theta_{1}}
\\
  &\ \ \ \ \ 
  \left[
    \sin\theta_{1}
    \left\{
      \left(
        \alpha_{1}
        \frac{\partial F}{\partial\theta_{1}}
        +
        \frac{1}{\sin\theta_{1}}
        \frac{\partial F}{\partial\varphi_{1}}
        +
        a_{J}M_{1}V_{1}
        \sin\theta_{1}
      \right)
      W
    \right.
  \right.
\\
  &\ \ \ \ \ \ \ \ \ \ \ \ \ \ \ \ \ \ \ \ \ \ \ \ 
      +
    \left.
  \left.
      \alpha_{1}
      k_{\rm B}T
      \frac{\partial W}{\partial\theta_{1}}
    \right\}
  \right]
\\
  &+
  \frac{\gamma_{1}}{M_{1}V_{1}}
  \frac{1}{\sin\theta_{1}}
  \frac{\partial}{\partial\varphi_{1}}
\\
  &\ \ \ \ \ 
  \left[
    \left(
      \frac{\alpha_{1}}{\sin\theta_{1}}
      \frac{\partial F}{\partial\varphi_{1}}
      -
      \frac{\partial F}{\partial\theta_{1}}
    \right)
    W
    +
    \frac{\alpha_{1}k_{\rm B}T}{\sin\theta_{1}}
    \frac{\partial W}{\partial\varphi_{1}}
  \right]
\\
  &+
  \frac{\gamma_{2}}{M_{2}V_{2}}
  \frac{1}{\sin\theta_{2}}
  \frac{\partial}{\partial\theta_{2}}
\\
  &\ \ \ \ \ 
  \left[
    \sin\theta_{2}
    \left\{
      \left(
        \alpha_{2}
        \frac{\partial F}{\partial\theta_{2}}
        +
        \frac{1}{\sin\theta_{2}}
        \frac{\partial F}{\partial\varphi_{2}}
      \right)
      W
      +
      \alpha_{2}
      k_{\rm B}T
      \frac{\partial W}{\partial\theta_{2}}
    \right\}
  \right]
\\
  &+
  \frac{\gamma_{2}}{M_{2}V_{2}}
  \frac{1}{\sin\theta_{2}}
  \frac{\partial}{\partial\varphi_{2}}
\\
  &\ \ \ \ \ 
  \left[
    \left(
      \frac{\alpha_{2}}{\sin\theta_{2}}
      \frac{\partial F}{\partial\varphi_{2}}
      -
      \frac{\partial F}{\partial\theta_{2}}
    \right)
    W
    +
    \frac{\alpha_{2}k_{\rm B}T}{\sin\theta_{2}}
    \frac{\partial W}{\partial\varphi_{2}}
  \right]\ .
  \label{eq:Fokker_Planck}
\end{split}
\end{equation}
Here we approximate that 
$1+\alpha_{k}^{2}\simeq 1$ 
by assuming that 
$\alpha_{k} \ll 1$ \cite{oogane06}. 
We also neglect the term proportional to $\alpha a_{J}$
by assuming that $|a_{J}|<|\mathbf{H}_{k}|$, 
which is valid in 
the thermally assisted switching region. 

As shown by Brown \cite{brown63}, 
the switching rate 
of the single ferromagnetic layer 
without spin transfer torque 
can be derived 
by using the steady-state solution 
of the Fokker-Planck equation 
and the continuity equation 
of the particles of an ensemble 
[see Sec. 4. C in Ref. \cite{brown63}]. 
In the case of two ferromagnetic layers,
as considered in this paper, 
the switching is 
described by the particle flow 
in $(\theta_{1},\varphi_{1},\theta_{2},\varphi_{2})$ 
four-dimensional phase space, 
and, in general, 
it is very difficult to obtain 
an analytical expression of 
the switching rate 
because the particle flow 
in the phase space is very complicated. 
To simplify the problem, 
we use the following two approximations.


\begin{figure}
  \centerline{\includegraphics[width=1.0\columnwidth]{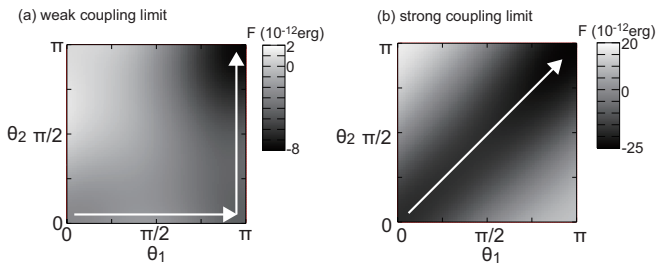}}
  \caption{
           The dependences of the effective potential $\mathscr{F}(\theta_{1},\theta_{2})$
           on $\theta_{1}$ and $\theta_{2}$ 
           for (a) the weak coupling limit and (b) the strong coupling limit. 
           The values of the parameters are written in 
           Secs. \ref{sec:Weak coupling limit} and \ref{sec:Strong coupling limit}.
           The white arrows indicate 
           the most probable paths of the switchings 
           $\mathbf{m}_{1},\mathbf{m}_{2}=+\mathbf{e}_{z}\to-\mathbf{e}_{z}$. 
  }
  \label{fig:fig2}
\end{figure}


First, 
we assume that 
the magnetization rotates
in the $yz$ plane 
during the switching. 
Since the deviation of the magnetization $\mathbf{m}_{k}$
from the $yz$ plane increases 
the magnetic energy due to the demagnetization field, 
it is reasonable to assume that 
the most probable reversal process 
is the magnetization reversal 
in the $yz$ plane. 
In this limit, 
the demagnetization field plays no role 
on the calculation of 
the switching probability. 
By fixing the values of $\varphi_{1}$ and $\varphi_{2}$ to 
$\pi/2$ or $3\pi/2$, 
the steady-state solution of the Fokker-Planck equation (\ref{eq:Fokker_Planck}) 
is given by $W_{0}\propto \exp[-\mathscr{F}/(k_{\rm B}T)]$, 
where the effective free energy $\mathscr{F}$ is given by 
\begin{equation}
  \mathscr{F}
  =
  F
  -
  \frac{a_{J}M_{1}V_{1}}{\alpha_{1}}
  \cos\theta_{1}.
  \label{eq:effective_free_energy}
\end{equation}
The switching probability is calculated 
by using $W_{0}\propto \exp[-\mathscr{F}/(k_{\rm B}T)]$. 

The second approximation is that 
we consider the switching 
for the weak and strong coupling limits, 
where the weak (strong) coupling means that 
the magnitude of 
the coupling energy of the F${}_{1}$ and F${}_{2}$ layers, 
$|-JS\mathbf{m}_{1}\cdot\mathbf{m}_{2}|$, 
is much smaller (larger) than 
the uniaxial anisotropy energy $M_{k}H_{{\rm an}k}V_{k}/2$. 
In other words, 
the weak (strong) coupling limit corresponds to 
$H_{{\rm an}k} \gg (\ll) H_{Jk}$. 
The weak and strong coupling limit can be realized by 
changing the thickness of the nonmagnetic layer $d_{\rm N}$ 
between F${}_{1}$ and F${}_{2}$ layers 
because the magnitude of the coupling constant $J$ strongly depends on $d_{\rm N}$
[see, for example, Fig. 2 in Ref. \cite{hayakawa08} or Fig. 1 (c) in Ref. \cite{yakata10}], 
and varies from $H_{J}\sim 1 \times 10^{3}$ Oe to $1$ Oe or less. 
Figures \ref{fig:fig2} (a) and (b) show
the dependences of $\mathscr{F}$ on $(\theta_{1},\theta_{2})$ 
for the weak and the strong coupling limit, respectively, 
where the white arrows indicate the most probable paths 
of the switching 
(the values of the parameters are written in 
Secs. \ref{sec:Weak coupling limit} and \ref{sec:Strong coupling limit}
with $a_{J}/a_{\rm c1}=0.5$). 
In the weak coupling limit, 
the magnetization reversal is divided into two steps: 
First $\mathbf{m}_{1}$ reverses 
its direction from $\mathbf{m}_{1}=+\mathbf{e}_{z}$ to 
$\mathbf{m}_{2}=-\mathbf{e}_{z}$ 
by the thermally assisted spin transfer torque effect 
while the direction of $\mathbf{m}_{2}$ is fixed 
to $\mathbf{m}_{2}=+\mathbf{e}_{z}$, 
and second, $\mathbf{m}_{2}$ reverses 
its direction 
by the thermal effect 
and the coupling with the F${}_{1}$ layer 
while $\mathbf{m}_{1}$ is fixed to $\mathbf{m}_{1}=-\mathbf{e}_{z}$. 
On the other hand, 
in the strong coupling limit, 
$\mathbf{m}_{1}$ and $\mathbf{m}_{2}$ reverse 
their directions simultaneously. 

By using the above two approximations, 
the calculation of the switching rate is 
reduced to a one-dimensional problem. 
For the weak coupling limit, 
first we calculate the particle flow 
in $\theta_{1}$ space, 
and second, 
we calculate the particle flow 
in $\theta_{2}$ space. 
On the other hand, 
for the strong coupling limit, 
we calculate the particle flow 
along the direction of $\theta_{1}=\theta_{2}$. 
For such one dimensional problems, 
the calculation method 
developed by Brown \cite{brown63} is applicable 
to obtain the switching rate 
with some revisions. 
In Secs. \ref{sec:Weak coupling limit} and \ref{sec:Strong coupling limit}, 
we show the switching probabilities 
for the weak and strong coupling limit, 
respectively. 

At the end of this section, 
we give a brief comment on the first approximation. 
The influence of the first approximation is that 
the critical current density estimated in our calculation, 
$I_{\rm c}\propto (H_{\rm appl}+H_{\rm an})$,
does not include the effect of the demagnetization field $4\pi M$
[see Eqs. (\ref{eq:critical_field_weak_1}), (\ref{eq:critical_field_weak_2}), 
(\ref{eq:critical_field_strong_1}) and (\ref{eq:critical_field_strong_2})] 
while the critical current density estimated by the LLG equation 
includes the demagnetization field, 
that is, $I_{\rm c}\propto (H_{\rm appl}+H_{\rm an}+2\pi M)$ 
[see, for example, Eq. (14) in Ref. \cite{li04}]. 
Since $4\pi M \gg |H_{\rm appl}|,H_{\rm an}$, 
the critical current density in our formula ($10^{5-6}$ A/cm${}^{2}$) is 
much smaller than the experimental values 
($10^{6-7}$ A/cm${}^{2}$)
\cite{hayakawa08,yakata09,yakata10}. 
One way to solve this discrepancy is as follows. 
Although we consider the in-plane magnetized system, 
it should be noted that our calculation is directly applicable 
to the perpendicularly magnetized system 
where the system has uni-axial symmetry 
and the switchings in the weak and strong coupling limits are 
described by only $\theta_{1}$ and $\theta_{2}$. 
Suzuki \etal \cite{suzuki09} showed that 
the effect of the demagnetization field 
on the switching rate of the in-plane magnetized system 
can be taken into account by replacing 
$H_{\rm an}$ in the switching formula of the perpendicularly magnetized system 
by $H_{\rm an}+2\pi M$. 
By applying this replacement to our formula, 
our formula may be applicable to 
analyze the experiments quantitatively. 
The validity of this replacement requires the numerical calculation 
of the Fokker-Planck equation, 
and it is beyond the scope of this paper.






\section{Weak coupling limit}
\label{sec:Weak coupling limit}

In this section, 
we derive the switching rate of 
the magnetizations 
for the weak coupling limit 
($H_{{\rm an}k}\gg H_{Jk}$)
[see also the Appendix]. 
With this limit, 
the magnetization reversal 
is divided into two steps, 
as mentioned in 
Sec. \ref{sec:Fokker-Planck equation for synthetic free layer}. 
For convenience, 
we label the three regions around
the potential minimum in the phase space, 
$(\theta_{1},\theta_{2})=(0,0),(\pi,0)$, and $(\pi,\pi)$, 
as regions 1, 2, and 3, respectively. 
The first step 
($\mathbf{m}_{1}$ reverses from $+\mathbf{e}_{z}$ to $-\mathbf{e}_{z}$) 
corresponds to the transition of the particle 
from region 1 to region 2 
while the second step 
($\mathbf{m}_{2}$ reverses from $+\mathbf{e}_{z}$ to $-\mathbf{e}_{z}$)
corresponds to the transition 
from region 2 to region 3. 

The switching rate 
from region 1 to region 2 
is obtained as follows \cite{brown63}. 
In regions 1 and 2, 
the distribution $W(\theta_{1},\theta_{2})$ is given by 
$W_{1}\exp[-\{\mathscr{F}(\theta_{1},0)-\mathscr{F}(0,0)\}/(k_{\rm B}T)]$
and $W_{2}\exp[-\{\mathscr{F}(\theta_{1},0)-\mathscr{F}(\pi,0)\}/(k_{\rm B}T)]$, 
respectively, where $W_{1}=W(0,0)$ and $W_{2}=W(\pi,0)$. 
The numbers of particles in region 1, $n_{1}$, 
is obtained by integrating 
$W(\theta_{1},0)$ over $[0,\theta_{\rm m1}]$, 
where $\theta_{\rm m1}=\cos^{-1}[-(H_{\rm appl}+H_{J1}+a_{J}/\alpha_{1})/H_{\rm an1}]$ 
gives the local maximum of the effective potential $\mathscr{F}(\theta_{1},0)$. 
The explicit form of $n_{1}$ is given by 
$n_{1}=2W_{1}{\rm e}^{\mathscr{F}(0,0)/(k_{\rm B}T)}I_{1}$, 
where factor 2 arises from the fact that 
we restrict the particle flow 
in the $yz$ plane;
that is, $\varphi_{1}=\pi/2$ or $3\pi/2$
(in the anisotropic system considered by Brown \cite{brown63}, 
the numerical factor is $2\pi$, not $2$, 
as shown in Eq. (4.26) of Ref. \cite{brown63}). 
The integral 
$I_{1}=\int_{0}^{\theta_{\rm m1}}{\rm d}\theta_{1} \sin\theta_{1} \exp[-\mathscr{F}(\theta_{1},0)/(k_{\rm B}T)]$ 
can be approximated to \cite{brown63}
\begin{equation}
\begin{split}
  I_{1}
  &\simeq 
  {\rm e}^{-\mathscr{F}(0,0)/(k_{\rm B}T)}
  \int_{0}^{\infty} {\rm d}\theta_{1}
  \theta_{1}
  \exp
  \left[
    -\frac{1}{2k_{\rm B}T}
    \frac{\partial^{2}\mathscr{F}(\theta_{1},0)}{\partial\theta_{1}^{2}}
    \theta_{1}^{2}
  \right]
\\
  &=
  {\rm e}^{-\mathscr{F}(0,0)/(k_{\rm B}T)}
  \frac{k_{\rm B}T}{\partial^{2}\mathscr{F}(0,0)/\partial\theta_{1}^{2}}.
  \label{eq:I_1}
\end{split}
\end{equation}
The numbers of particle in region 2, 
$n_{2}=2W_{2}{\rm e}^{\mathscr{F}(\pi,0)/(k_{\rm B}T)}I_{2}$, 
is obtained in a similar way 
by replacing the factors 
$\mathscr{F}(0,0)$ and $\partial^{2}\mathscr{F}(0,0)/\partial\theta_{1}^{2}$ 
to $\mathscr{F}(\pi,0)$ and 
$\partial^{2}\mathscr{F}(\pi,0)/\partial\theta_{1}^{2}$, 
respectively. 
Next, we consider the particle flow 
from region 1 to region 2, $I_{1 \to 2}$. 
From the Fokker-Planck equation (\ref{eq:Fokker_Planck}), 
the particle flow along the $\theta_{1}$-axis, $J_{\theta_{1}}$, 
which satisfies $I_{1 \to 2}=2 \sin\theta_{1} J_{\theta_{1}}$, 
is identified as 
\begin{equation}
  J_{\theta_{1}}
  =
  -\frac{\alpha_{1}\gamma_{1}}{M_{1}V_{1}}
  \left[
    \left(
      \frac{\partial F}{\partial\theta_{1}}
      +
      \frac{a_{J}M_{1}V_{1}}{\alpha_{1}}
      \sin\theta_{1}
    \right)
    W
    +
    k_{\rm B}T 
    \frac{\partial W}{\partial \theta_{1}}
  \right]. 
  \label{eq:J_theta1}
\end{equation}
By multiplying ${\rm e}^{\mathscr{F}(\theta_{1},0)/(k_{\rm B}T)}$ 
to $I_{1 \to 2}/(2 \sin\theta_{1})=J_{\theta_{1}}$ 
and integrating it over $[0,\pi]$, 
we find that 
$[(n_{2}/I_{2})-(n_{1}/I_{1})]/2=-[M_{1}V_{1}/(2\alpha_{1}\gamma_{1}k_{\rm B}T)]I_{1 \to 2}I_{\rm m1}$, 
where the integral 
$I_{\rm m1}=\int_{0}^{\pi} {\rm d}\theta_{1} {\rm e}^{\mathscr{F}(\theta_{1},0)/(k_{\rm B}T)}/\sin\theta_{1}$ 
can be approximated to \cite{brown63} 
\begin{equation}
\begin{split}
  I_{\rm m1}
  &\simeq 
  \frac{e^{\mathscr{F}(\theta_{\rm m1},0)/(k_{\rm B}T)}}{\sin\theta_{\rm m1}}
\\
  &\ \ \ \ \ \times
  \int_{-\infty}^{\infty} {\rm d}\theta_{1}
  \exp
  \left[
    \frac{(\theta_{1}-\theta_{\rm m1})^{2}}{2k_{\rm B}T}
    \frac{\partial^{2}\mathscr{F}(\theta_{\rm m1},0)}{\partial\theta_{1}^{2}}
  \right]
\\
  &=
  \sqrt{
    -\frac{2\pi k_{\rm B}T}{\partial^{2}\mathscr{F}(\theta_{\rm m1},0)/\partial\theta_{1}^{2}}
  }
  \frac{{\rm e}^{\mathscr{F}(\theta_{\rm m1},0)/(k_{\rm B}T)}}{\sin\theta_{\rm m1}}.
  \label{eq:I_m1}
\end{split}
\end{equation}
The relation between 
the particle numbers in region 2 and 3, 
$n_{2}$ and $n_{3}$, 
and the particle flow 
from region 2 to region 3, $I_{2 \to 3}$, 
is obtained in a similar way. 
Then, by using the continuity equations of the particle flow, 
$\dot{n}_{1}=-I_{1 \to 2}$, 
$\dot{n}_{2}=I_{1 \to 2}-I_{2 \to 3}$, 
and $\dot{n}_{3}=I_{2 \to 3}$, 
we find that 
the transitions of the magnetization directions 
among the three states, 
$(\theta_{1},\theta_{2})=(0,0),(\pi,0),(\pi,\pi)$,
are described by the following differential equations: 
\begin{equation}
  \frac{{\rm d}}{{\rm d}t}
  \begin{pmatrix}
    n_{1} \\
    n_{2} \\
    n_{3}
  \end{pmatrix}
  =
  \begin{pmatrix}
    -\nu_{12} & \nu_{21} & 0 \\
    \nu_{12} & -(\nu_{21}+\nu_{23}) & \nu_{32} \\
    0 & \nu_{23} & -\nu_{32} 
  \end{pmatrix}
  \begin{pmatrix}
    n_{1} \\
    n_{2} \\
    n_{3}
  \end{pmatrix}. 
  \label{eq:transition_equation_weak_limit}
\end{equation}
The switching probability per unit time 
from the region $i$ to the region $j$ is given by 
$\nu_{ij}=f_{ij}\exp(-\Delta_{ij})$, 
where the attempt frequency $f_{ij}$
and the thermal stability $\Delta_{ij}$ are, 
respectively, given by 
\begin{equation}
\begin{split}
  f_{12(21)}
  &=
  \left(
    \frac{\alpha_{1}\gamma_{1}k_{\rm B}T}{M_{1}V_{1}}
  \right)
  \frac{M_{1}H_{\rm an1}V_{1}}{k_{\rm B}T}
  \sqrt{
    \frac{M_{1}H_{\rm an1}V_{1}}{2\pi k_{\rm B}T}
  }
\\
  &\ \times
  \left(
    1
    +(-)
    \frac{H_{\rm appl}+H_{J1}}{H_{\rm an1}}
  \right)
  \left(
    1
    -
    \frac{I}{I_{\rm c1(2)}}
  \right)
\\
  &\ \times
  \left[
    1
    -
    \left(
      \frac{H_{\rm appl}+H_{J1}}{H_{\rm an1}}
    \right)^{2}
  \right]
  \left(
    1
    -
    \frac{I}{I_{\rm c1}}
  \right)
  \left(
    1
    -
    \frac{I}{I_{\rm c2}}
  \right),
  \label{eq:f_12_weak}
\end{split}
\end{equation}
\begin{equation}
\begin{split}
  \Delta_{12(21)}
  &=
  \frac{M_{1}H_{\rm an1}V_{1}}{2k_{\rm B}T}
  \left(
    1
    +(-)
    \frac{H_{\rm appl}+H_{J1}}{H_{\rm an1}}
  \right)^{2}
\\
  &\ \times
  \left(
    1
    -
    \frac{I}{I_{\rm c1(2)}}
  \right)^{2},
  \label{eq:delta_weak}
\end{split}
\end{equation}
\begin{equation}
\begin{split}
  f_{23(32)}
  &=
  \left(
    \frac{\alpha_{2}\gamma_{2}k_{\rm B}T}{M_{2}V_{2}}
  \right)
  \frac{M_{2}H_{\rm an2}V_{2}}{k_{\rm B}T}
  \sqrt{
    \frac{M_{2}H_{\rm an2}V_{2}}{2\pi k_{\rm B}T}
  }
\\
  &\ \times
  \left(
    1
    +(-)
    \frac{H_{\rm appl}-H_{J2}}{H_{\rm an2}}
  \right)
\\
  &\ \times
  \left[
    1
    -
    \left(
      \frac{H_{\rm appl}-H_{J2}}{H_{\rm an2}}
    \right)^{2}
  \right],
  \label{eq:f_23_weak}
\end{split}
\end{equation}
\begin{equation}
  \Delta_{23(32)}
  =
  \frac{M_{2}H_{\rm an2}V_{2}}{2k_{\rm B}T}
  \left(
    1
    +(-)
    \frac{H_{\rm appl}-H_{J2}}{H_{\rm an2}}
  \right)^{2}.
  \label{eq:delta_weak_sub}
\end{equation}
Here $I/I_{\rm c1}=a_{J}/a_{\rm c1}$ and 
$I/I_{\rm c2}=a_{J}/a_{\rm c2}$. 
$a_{\rm c1}$ ($a_{\rm c2}$) is 
the critical spin-transfer torque field 
to induce the magnetization reversal 
from region 1 (2) to region 2 (1) 
at zero temperature, 
and their explicit forms are given by 
\begin{equation}
  a_{\rm c1}
  =
  -\alpha_{1}
  \left(
    H_{\rm appl}
    +
    H_{J1}
    +
    H_{\rm an1}
  \right),
  \label{eq:critical_field_weak_1}
\end{equation}
\begin{equation}
  a_{\rm c2}
  =
  \alpha_{1}
  \left(
    -H_{\rm appl}
    -
    H_{J1}
    +
    H_{\rm an1}
  \right),
  \label{eq:critical_field_weak_2}
\end{equation}
respectively. 
Since $|H_{\rm appl}+H_{J1}|$ is 
assumed to be smaller than $H_{\rm an1}$, 
we find that $a_{\rm c1}<0$ and $a_{\rm c2}>0$. 
It should be noted that 
the description of the transition of the magnetization 
by Eq. (\ref{eq:transition_equation_weak_limit}) 
is valid for $|a_{J}| < |a_{{\rm c}k}|$ 
because if $|a_{J}| \gg |a_{\rm c1}| (|a_{\rm c2}|)$, 
the point $\mathbf{m}_{1}=+\mathbf{e}_{z} (-\mathbf{e}_{z})$ would be unstable, 
and then 
we could not discuss 
the thermally assisted transition. 
We also note that
the switching probabilities of $\mathbf{m}_{2}$, 
$\nu_{23}$ and $\nu_{32}$, 
are reduced to those obtained by Brown \cite{brown63} 
by omitting $H_{J2}$ 
where $\nu_{23}$ and $\nu_{32}$ are 
independent of the current $I$. 


\begin{figure}
  \centerline{\includegraphics[width=0.8\columnwidth]{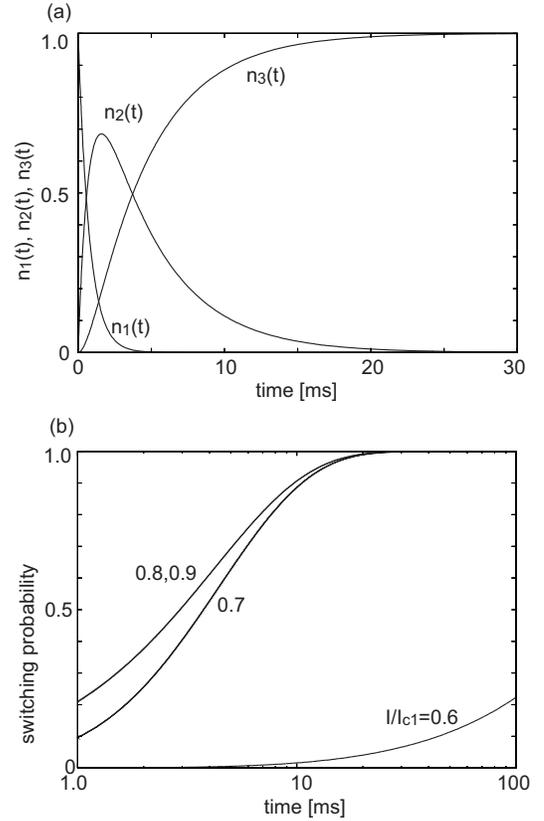}}
  \caption{
           (a) Time evolution of $n_{1}(t)$, $n_{2}(t)$, and $n_{3}(t)$ 
               with $I/I_{\rm c1}=0.7$ 
               for the weak coupling limit. 
           (b) The dependence of the switching rate $n_{3}(t)$ on the ratio $I/I_{\rm c1}$
               for the weak coupling limit.
               For $I/I_{\rm c1} \ge 0.8$, 
               the switching time is saturated. 
               The horizontal axis is the logarithmic scale. 
  }
  \label{fig:fig3}
\end{figure}


When $I$ is nearly $I_{\rm c1}$, 
we find that 
$\nu_{12}/\nu_{21}\sim \exp[M_{1}H_{\rm an1}V_{1}/(2k_{\rm B}T)] \gg 1$. 
Similarly, when $-H_{\rm appl}+H_{J2}>0$, 
we find that $\nu_{23}/\nu_{32}\sim \exp[2M_{2}(-H_{\rm appl}+H_{J2})V_{2}/(2k_{\rm B}T)]\gg 1$. 
Within these limits, 
the analytical solutions 
of Eq. (\ref{eq:transition_equation_weak_limit}) 
with the initial conditions 
$n_{1}(0)=1$, $n_{2}(0)=0$, and $n_{3}(0)=0$, 
are given by 
\begin{equation}
  n_{1}(t)
  =
  {\rm e}^{-\nu_{12}t},
  \label{eq:n1_weak}
\end{equation}
\begin{equation}
  n_{2}(t)
  =
  -\frac{\nu_{12}}{\nu_{12}-\nu_{23}}
  \left(
    {\rm e}^{-\nu_{12}t}
    -
    {\rm e}^{-\nu_{23}t}
  \right), 
  \label{eq:n2_weak}
\end{equation}
\begin{equation}
  n_{3}(t)
  =
  1
  -
  \frac{\nu_{12}{\rm e}^{-\nu_{23}t}-\nu_{23}{\rm e}^{-\nu_{12}t}}{\nu_{12}-\nu_{23}}.
  \label{eq:switching_rate_weak}
\end{equation}
Equation (\ref{eq:switching_rate_weak}) is 
the central result of this section: 
It completely describes 
the magnetization switching of the synthetic free layer 
within the weak coupling limit. 

Figure \ref{fig:fig3} (a) shows 
a typical time evolution of 
$n_{1}(t)$, $n_{2}(t)$, and $n_{3}(t)$ 
for a synthetic free layer with 
$M=995$ emu/c.c., 
$H_{\rm an}=50$ Oe,
$H_{\rm appl}=0$ Oe,
$\alpha=0.007$, 
$\gamma=1.732\times 10^{7}$ Hz/Oe, 
$d=2$ nm, 
$S=\pi\times 70 \times 160$ nm${}^{2}$, 
and $T=300$ K
(for simplicity, we assume that F${}_{1}$=F${}_{2}$) 
\cite{yakata09,yakata10,comment3}. 
The current is taken to be $I/I_{\rm c1}=0.7$. 
The coupling constant is assumed to be 
$J=5.0\times 10^{-3}$ erg/cm${}^{2}$, 
which corresponds to $H_{J}=25$ Oe. 
From Eqs. (\ref{eq:n1_weak}), (\ref{eq:n2_weak}) and (\ref{eq:switching_rate_weak}), 
one can easily see that 
the time evolution shown in Fig. \ref{fig:fig3} (a) 
is determined by two time scales, 
$\nu_{12}^{-1}$ and $\nu_{23}^{-1}$, 
which correspond to the switching rates of
the F${}_{1}$ and F${}_{2}$ layers, respectively. 
Figure \ref{fig:fig3} (b) shows 
the dependence of the switching rate $n_{3}(t)$ 
on the ratio $I/I_{\rm c1}$. 
For the currents $|I| \ge 0.7 |I_{\rm c1}|$, 
the switching times are on the same order 
($10$ ms for our parameters), 
and for the large currents $|I| \ge 0.8 |I_{\rm c1}|$, 
the switching times are saturated. 
This is because 
the current determines the switching time 
of the F${}_{1}$ layer only, 
and for a large current, 
the total switching time of $\mathbf{m}_{1}$ and $\mathbf{m}_{2}$ 
is mainly determined by 
the switching time of $\mathbf{m}_{2}$, 
which is independent of the current. 
We can verify the saturation of the switching time 
from Eq. (\ref{eq:switching_rate_weak}), 
where $\nu_{12}$ becomes much larger than $\nu_{23}$ 
as $I$ approaches $I_{\rm c1}$
and then, $n_{3}(t)\simeq 1-{\rm e}^{-\nu_{23}t}$, 
which is independent of the current $I$. 
On the other hand, 
in the low current region $|I| \ll |I_{\rm c1}|$, 
$\nu_{12}$ becomes comparable or smaller than $\nu_{23}$, 
which leads to $n_{3}(t)\simeq 1-{\rm e}^{-\nu_{12}t}$. 
Then the switching time strongly depends on the current value 
because the switching time of $\mathbf{m}_{1}$ 
becomes important to the total switching time. 
For example, 
the switching time for $I/I_{\rm c1}=0.6$ is 
longer than 100 ms, 
as shown in Fig. \ref{fig:fig3} (b). 

The dependence of the switching time on the coupling constant $J$ is as follow.
By increasing the magnitude of $H_{J}$, 
the switching time rapidly decreases 
because of the fast reversal of the magentization 
of the F${}_{2}$ layer. 
For example, for $H_{J}=40$ Oe with $I/I_{\rm c}=0.9$, 
the switching time is on the order of $10^{-2}$ ms, 
which is three orders of magnitude faster than 
that for $H_{J}=25$ Oe. 
On the other hand, by decreasing the magnitude of $H_{J}$, 
only the magnetization of the F${}_{1}$ layer 
reverses its direction 
while the magnetization of the F${}_{2}$ layer 
remains $\mathbf{m}_{2}=+\mathbf{e}_{z}$. 
For example, for $H_{J}=5$ Oe with $I/I_{\rm c}=0.9$, 
the switching time of the F${}_{1}$ layer is 
on the order of $10^{-2}$ ms 
while the switching rate of the F${}_{2}$ layer, $n_{3}$, 
is approximately zero ($n_{3}\sim 10^{-9}$). 
Since the switching of the F${}_{2}$ layer is 
induced by the coupling with the F${}_{1}$ layer, 
it is required to increase the magnitude of the coupling constant $J$
for the fast switching 
by using a thin nonmagnetic spacer, 
although the increase of the coupling constant leads 
to the increase of the magnitude of the critical current density. 



\begin{figure}
  \centerline{\includegraphics[width=0.8\columnwidth]{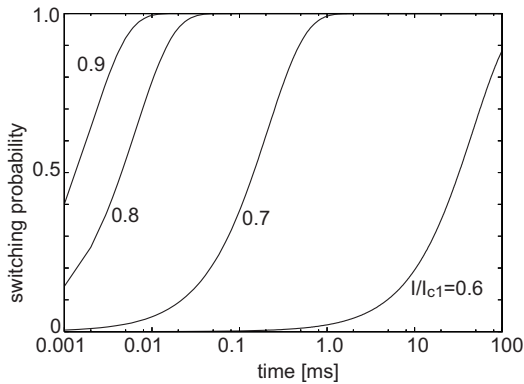}}
  \caption{
           The dependence of the switching rate $n_{2}(t)$ on the ratio $I/I_{\rm c1}$ 
           for the strong coupling limit.
           The horizontal axis is the logarithmic scale.
  }
  \label{fig:fig4}
\end{figure}



\section{Strong coupling limit}
\label{sec:Strong coupling limit}

In this section, 
we derive the switching rate of the magnetizations 
for the strong coupling limit ($H_{{\rm an}k} \ll H_{Jk}$). 
For this limit, 
instead of $(\theta_{1},\theta_{2})$ phase space, 
it is convenient to describe the particle flow 
in $(\Psi,\psi)$ phase space, 
where $\Psi=(\theta_{1}+\theta_{2})/2$ and 
$\psi=\theta_{1}-\theta_{2}$. 
Since $\mathbf{m}_{1}$ and $\mathbf{m}_{2}$ 
reverse their directions simultaneously, 
the reversal is described by 
the particle flow along the $\Psi$-axis 
with $\psi=0$.
For convenience, 
we label the two regions around 
the potential minimum in the phase space, 
$(\Psi,\psi)=(0,0)$ and $(\pi,0)$, 
as regions 1 and 2, respectively. 
The continuity equation of 
the particle in the regions 1 and 2 is 
obtained in a way similar to that 
described in Sec. \ref{sec:Weak coupling limit} 
and is expressed as 
$\dot{n}_{1}=-\dot{n}_{2}=-\nu_{12}n_{1}+\nu_{21}n_{2}$.
The switching probability $\nu_{ij}=f_{ij}\exp(-\Delta_{ij})$ is given by 
\begin{equation}
\begin{split}
  f_{12(21)}
  &=
  \frac{k_{\rm B}T}{2}
  \left(
    \frac{\alpha_{1}\gamma_{1}}{M_{1}V_{1}}
    +
    \frac{\alpha_{2}\gamma_{2}}{M_{2}V_{2}}
  \right)
  \frac{M_{1}H_{\rm an1}V_{1}+M_{2}H_{\rm an2}V_{2}}{k_{\rm B}T}
\\
  &\ \times
  \sqrt{
    \frac{M_{1}H_{\rm an1}V_{1}+M_{2}H_{\rm an2}V_{2}}{2\pi k_{\rm B}T}
  }
\\
  &\ \times
  \left(
    1
    +(-)
    \frac{M_{1}H_{\rm appl}V_{1}+M_{2}H_{\rm appl}V_{2}}{M_{1}H_{\rm an1}V_{1}+M_{2}H_{\rm an2}V_{2}}
  \right)
  \left(
    1
    -
    \frac{I}{I_{\rm c1(2)}}
  \right)
\\
  &\ \times
  \left[
    1
    -
    \left(
      \frac{M_{1}H_{\rm appl}V_{1}+M_{2}H_{\rm appl}V_{2}}{M_{1}H_{\rm an1}V_{1}+M_{2}H_{\rm an2}V_{2}}
    \right)^{2}
  \right]
\\
  &\ \times
  \left(
    1
    -
    \frac{I}{I_{\rm c1}}
  \right)
  \left(
    1
    -
    \frac{I}{I_{\rm c2}}
  \right),
\end{split}
\end{equation}
\begin{equation}
\begin{split}
  \Delta_{12(21)}
  &=
  \frac{M_{1}H_{\rm an1}V_{1}+M_{2}H_{\rm an2}V_{2}}{2k_{\rm B}T}
\\
  &\ \times
  \left(
    1
    +(-)
    \frac{M_{1}H_{\rm appl}V_{1}+M_{2}H_{\rm appl}V_{2}}{M_{1}H_{\rm an1}V_{1}+M_{2}H_{\rm an2}V_{2}}
  \right)^{2}
\\
  &\ \times
  \left(
    1
    -
    \frac{I}{I_{\rm c1(2)}}
  \right)^{2}, 
  \label{eq:delta_strong}
\end{split}
\end{equation}
where $I/I_{\rm c1}=a_{J}/a_{\rm c1}$ and $I/I_{\rm c2}=a_{J}/a_{\rm c2}$, 
and the critical spin-transfer torque fields 
in the strong coupling limit $a_{{\rm c}k}$ are given by 
\begin{equation}
  a_{\rm c1}
  =
  -\alpha_{1}
  \left[
    H_{\rm appl}
    +
    H_{\rm an1}
    +
    \frac{M_{2}V_{2}}{M_{1}V_{1}}
    \left(
      H_{\rm appl}
      +
      H_{\rm an2}
    \right)
  \right],
  \label{eq:critical_field_strong_1}
\end{equation}
\begin{equation}
  a_{\rm c2}
  =
  \alpha_{1}
  \left[
    -H_{\rm appl}
    +
    H_{\rm an1}
    +
    \frac{M_{2}V_{2}}{M_{1}V_{1}}
    \left(
      -H_{\rm appl}
      +
      H_{\rm an2}
    \right)
  \right],
  \label{eq:critical_field_strong_2}
\end{equation}
respectively. 
The analytical solutions 
of the transition equations,
$\dot{n}_{1}=-\dot{n}_{2}=-\nu_{12}n_{1}+\nu_{21}n_{2}$, 
with the initial conditions 
$n_{1}(0)=1$ and $n_{2}(0)=0$ 
are given by 
\begin{equation}
  n_{1}(t)
  =
  \frac{\nu_{21}}{\nu_{12}+\nu_{21}}
  +
  \frac{\nu_{12}}{\nu_{12}+\nu_{21}}
  {\rm e}^{-(\nu_{12}+\nu_{21})t}, 
  \label{eq:n1_strong}
\end{equation}
\begin{equation}
  n_{2}(t)
  =
  \frac{\nu_{12}}{\nu_{12}+\nu_{21}}
  -
  \frac{\nu_{12}}{\nu_{12}+\nu_{21}}
  {\rm e}^{-(\nu_{12}+\nu_{21})t}. 
  \label{eq:n2_strong}
\end{equation}
When $I$ is nearly $I_{\rm c1}$, 
$\nu_{12}/\nu_{21}\sim \exp[(M_{1}H_{\rm an1}V_{1}+M_{2}H_{\rm an2}V_{2})/(2k_{\rm B}T)] \gg 1$.  
For this limit, 
Eqs. (\ref{eq:n1_strong}) and (\ref{eq:n2_strong}) are reduced to 
\begin{equation}
  n_{1}(t)
  \simeq
  {\rm e}^{-\nu_{12}t},
\end{equation}
\begin{equation}
  n_{2}(t)
  \simeq 
  1
  -
  {\rm e}^{-\nu_{12}t}. 
  \label{eq:switching_rate_strong}
\end{equation}
Equation (\ref{eq:switching_rate_strong}) is 
the central result of this section: 
It completely describes 
the magnetization switching of the synthetic free layer 
within the strong coupling limit.

For the strong coupling limit, 
the switching time strongly depends on 
the current $I$
for all current region.  
Figure \ref{fig:fig4} shows 
the dependence of $n_{2}(t)$ on the ratio $I/I_{\rm c1}$, 
where the parameters used are 
same as those in Fig. \ref{fig:fig3} 
except $J$. 
The coupling constant $J$ is assumed to be $5.0\times 10^{-2}$ erg/cm${}^{-2}$, 
which corresponds to $H_{J}=250$ Oe. 
The orders of the switching times are 
$10^{-2}$ ms for $I/I_{\rm c1}= 0.8,0.9$, 
$1$ ms for $I/I_{\rm c1}=0.7$, 
and more than $100$ ms for $I/I_{\rm c1} \le 0.6$ 
in our parameter region, 
as shown in Fig. \ref{fig:fig4}. 
Such strong dependence of the switching time on the current 
arises from 
the thermal stability $\Delta_{12}$, 
which is proportional to $(1-I/I_{\rm c1})^{2}$, 
as shown in Eq. (\ref{eq:delta_strong}).


\section{Relation to other works}
\label{sec:Relation to other works}

In this section, 
we compare the results obtained in the previous sections 
to the other works \cite{hayakawa08,yakata09,yakata10,koch04}. 
The topics discussed here are
(1) the comparison of the switching time 
of the ferromagnetically (F) 
and the anti-ferromagnetically (AF) coupled 
synthetic free layers, 
and (2) the comparison of 
the dependence of the thermal stability to 
that obtained by Koch \etal \cite{koch04}. 


\begin{figure}
  \centerline{\includegraphics[width=1.0\columnwidth]{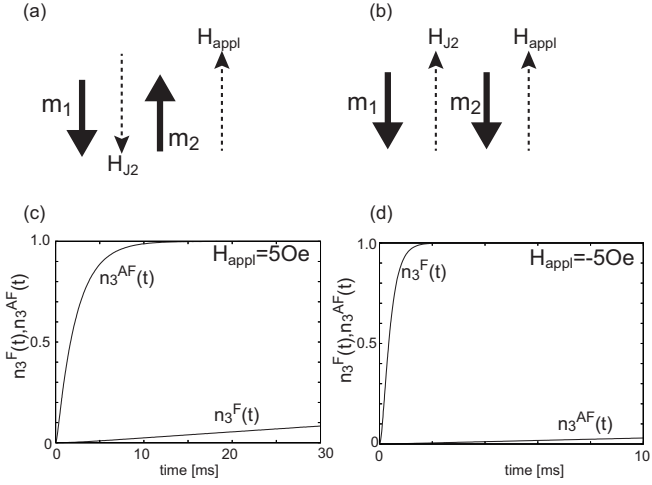}}
  \caption{
           (a), (b): 
           The schematic views of 
           the alignments of the magnetizations $\mathbf{m}_{1}$ and $\mathbf{m}_{2}$ 
           of (a) ferromagnetically (F) coupled 
           and (b) anti-ferromagnetically (AF) coupled synthetic free layers 
           after $\mathbf{m}_{1}$ reverses its direction 
           from the initial state ($\mathbf{m}_{1}=+\mathbf{e}_{z}$) to 
           $\mathbf{m}_{1}=-\mathbf{e}_{z}$. 
           The directions of the applied field $H_{\rm appl}$ (assumed to be positive) 
           and the coupling field $H_{J2}$ are also denoted.
           (c), (d): 
           The time evolution of $n_{3}(t)$ 
           of the F-coupled ($n_{3}^{({\rm F})}(t)$) and 
           the AF-coupled ($n_{3}^{({\rm AF})}(t)$) synthetic free layer 
           with (c) $H_{\rm appl}=+5$ Oe and (d) $H_{\rm appl}=-5$ Oe. 
           The value of the current is taken to be $I/I_{\rm c1}=0.7$. 
  }
  \label{fig:fig5}
\end{figure}


First, we discuss the switching times of 
the F and the AF-coupled 
synthetic free layers. 
The difference in the switching times of 
these two kinds of synthetic free layer appears 
in the weak coupling limit 
with finite $H_{\rm appl}$. 
In this case, 
the switching time of the F-coupled synthetic free layer 
is characterized by Eqs. (\ref{eq:delta_weak}) and (\ref{eq:delta_weak_sub}). 
For the AF-coupled synthetic free layer, 
the factor $+(-)(H_{\rm appl}-H_{J2})/H_{\rm an2}$ in Eq. (\ref{eq:delta_weak_sub}) 
is replaced by 
$-(+)(H_{\rm appl}+H_{J2})/H_{\rm an2}$ 
while Eq. (\ref{eq:delta_weak}) remains the same. 
This replacement is due to the fact that 
after $\mathbf{m}_{1}$ reverses its direction from $+\mathbf{e}_{z}$ to $-\mathbf{e}_{z}$, 
the sum of the applied field $H_{\rm appl}$ and the coupling field $H_{J2}$ 
acting on $\mathbf{m}_{2}$ is 
$H_{\rm appl}-H_{J2}$ for the F-coupled synthetic free layer 
while it is $H_{\rm appl}+H_{J2}$ for the AF-coupled synthetic free layer, 
as schematically shown in Figs. \ref{fig:fig5} (a) and (b), 
and leads to the difference in 
the switching times of the F-coupled and the AF-coupled 
synthetic layers.

The important point is that 
the fast switching is achieved by 
choosing the appropriate direction of $H_{\rm appl}$. 
Figures \ref{fig:fig5} (c) and (d) 
show the time evolutions of $n_{3}(t)$ (Eq. (\ref{eq:switching_rate_weak})) 
for the F-coupled ($n_{3}^{({\rm F})}$) and 
the AF-coupled ($n_{3}^{({\rm AF})}$) synthetic free layers 
with (c) $H_{\rm appl}=+5$ Oe and (d) $H_{\rm appl}=-5$ Oe. 
The current is taken to be $I/I_{\rm c1}=0.7$. 
The switching time of the AF (F) coupled synthetic free layer 
is faster compared to that of the F (AF) coupled synthetic free layer 
for $H_{\rm appl}>0(<0)$ 
because both $H_{\rm appl}$ and $H_{J2}$ assist 
the reversal of $\mathbf{m}_{2}$. 
On the other hand, 
by changing the direction (sign) of $H_{\rm appl}$, 
the switching time increases significantly 
because of the exponential dependence of 
the switching time ($\sim 1/\nu$) 
on $\Delta \propto [1-(H_{\rm appl}\pm H_{J})/H_{\rm an}]^{2}$. 
The difference between 
$n_{3}^{({\rm AF})}$ with $H_{\rm appl}>0$ and 
$n_{3}^{({\rm F})}$ with $H_{\rm appl}<0$ 
arises from the dependence of 
the switching time of $\mathbf{m}_{1}$ on the direction of $H_{\rm appl}$, 
and becomes negligible 
as $I$ approaches $I_{\rm c1}$ 
because the total switching time is mainly determined by 
that of $\mathbf{m}_{2}$ within the limit of $I/I_{\rm c1}\to 1$, 
as mentioned in Sec. \ref{sec:Weak coupling limit}. 
For the strong coupling limit, 
the switching times of the F-coupled and the AF-coupled 
synthetic free layers are the same 
because the coupling energy is constant 
during the switching in this limit, 
and the coupling field $H_{J}$ plays no role 
on the switching. 

Second, we discuss the dependence of 
the thermal stability $\Delta$ on the current $I$. 
As shown in Eqs. (\ref{eq:delta_weak}) and (\ref{eq:delta_strong}), 
our calculations show that 
$\Delta\propto \Delta_{0}(1-I/I_{\rm c})^{2}$. 
It should be noted that 
our formula is applicable to 
the single free layer by omitting 
the coupling of the F${}_{1}$ and the F${}_{2}$ layers,
and thus, 
even for the single free layer 
we find that 
$\Delta \propto \Delta_{0}(1-I/I_{\rm c})^{2}$. 
Recently, 
a similar result was obtained by Suzuki \etal \cite{suzuki09} 
and Butler \etal \cite{comment4} 
for the perpendicularly magnetized single free layer. 
However, 
the formula of 
the switching rate with $\Delta\propto \Delta_{0}(1-I/I_{\rm c})$ 
first obtained by Koch \etal \cite{koch04} 
has been widely used to fit the experiments \cite{hayakawa08,yakata09,yakata10}. 

The important point is that 
the difference of the exponent of $(1-I/I_{\rm c})$ leads to 
a significant underestimation of $\Delta_{0}$. 
Let us consider the fit of the experimental results of 
the switching rate 
with the formula 
$P=1-\exp\{-f_{0}t\exp[-\Delta_{0}(1-I/I_{\rm c})^{n}]\}$, 
where for simplicity we assume that 
the attempt frequency $f_{ij}$ is constant $f_{0}$. 
When $I/I_{\rm c}=0.5$, 
the thermal stability $\Delta_{0}$ estimated by our formula ($n=2$) 
is two times larger than 
that estimated by the conventional formula ($n=1$). 

The difference between the exponent of $(1-I/I_{\rm c})$ 
in our calculation and that in
the theory of Koch \etal \cite{koch04} 
arises from the steady-state solution of 
the Fokker-Planck equation of the free layer magnetization: 
\begin{equation}
\begin{split}
  \frac{\partial W}{\partial t}
  =&
  \frac{\alpha\gamma}{MV}
  \frac{1}{\sin\theta}
  \frac{\partial}{\partial\theta}
\\
  &\ \ 
  \left[
    \sin\theta
    \left\{
      \left(
        \frac{\partial F}{\partial\theta}
        +
        \frac{a_{J}MV}{\alpha}
        \sin\theta
      \right)
      W
      +
      k_{\rm B}T
      \frac{\partial W}{\partial\theta}
    \right\}
  \right].
  \label{eq:Fokker_Planck_Koch}
\end{split}
\end{equation}
Koch \etal argued that 
the steady-state solution of 
of Eq. (\ref{eq:Fokker_Planck_Koch}) is 
$W_{\rm Koch}\propto \exp\{-F[1+a_{J}/(\alpha H)]/(k_{\rm B}T)\}$, 
where $H=|\mathbf{H}|$ is 
the absolute value of the magnetic field 
acting on the free layer magnetization. 
However, when $H$ depends on $\mathbf{m}$, 
$W_{\rm Koch}$ is \textit{not} a steady state solution 
of Eq. (\ref{eq:Fokker_Planck_Koch}). 
In general, 
$H$ depends on $\mathbf{m}$ 
because of the presence of the uni-axial anisotropy field 
$\mathbf{H}_{\rm an}=H_{\rm an}m_{z}\mathbf{e}_{z}$, 
which guarantees two local minima of the free energy $F$. 
Thus, in the calculation of the switching rate, 
we should use 
$W_{0} \propto \exp\{-[1-(a_{J}MV\cos\theta)/(\alpha F)/(k_{\rm B}T)]\}$, 
which is the steady state solution of Eq. (\ref{eq:Fokker_Planck_Koch})
as shown in Sec. \ref{sec:Fokker-Planck equation for synthetic free layer}, 
instead of $W_{\rm Koch}$. 
The difference between $W_{0}$ and $W_{\rm Koch}$ 
leads to that of the exponent of $(1-I/I_{\rm c})$ in $\Delta$.


\section{Conclusions}
\label{sec:Conclusions}

In conclusion, 
we studied the magnetization switching 
of the synthetic free layer theoretically 
by solving the Fokker-Planck equation. 
We obtained the analytical expression 
of the switching rate 
for the weak and the strong coupling limits, 
given by Eqs. (\ref{eq:switching_rate_weak}) and (\ref{eq:switching_rate_strong}). 
We found that 
the switching time within the weak coupling limit becomes saturated 
as the current $I$ approaches the critical current $I_{\rm c1}$. 
We compared the switching time of 
the ferromagnetically and the anti-ferromagnetically 
coupled synthetic free layers 
with a finite applied field, 
and find that 
fast switching is achieved 
by choosing the appropriate direction of the applied field. 
We also found that 
the dependence of the thermal stability 
on the current is 
$\Delta\propto \Delta_{0}(1-I/I_{\rm c})^{2}$, 
not $\Delta\propto \Delta_{0}(1-I/I_{\rm c})$ 
as argued by previous authors \cite{koch04}, 
which leads to a significant underestimation 
of $\Delta_{0}$.





\section*{ACKNOWLEDGMENT}
\label{sec:Acknowledgment}

 The authors would like to acknowledge 
 H. Kubota, S. Yuasa, K. Seki, 
 M. Marthaler and D. S. Golubev 
 for valuable discussions. 
 This work was supported by JSPS and NEDO. 


\section*{APPENDIX A: DETAILS OF THE CALCULATION IN SEC. \ref{sec:Weak coupling limit}}
\label{eq:Appendix_A}

In this appendix, we show the details of 
the derivation of Eq. (\ref{eq:switching_rate_weak})
[see also Sec. 4 C in Ref. \cite{brown63}]. 
First, let us consider the switching from region 1 to region 2. 
The number of the particle in region 1 is obtained 
by integrating $2W_{1}\exp[-\{\mathscr{F}(\theta_{1},0)-\mathscr{F}(0,0)\}/(k_{\rm B}T)]$ 
over $0 \le \theta_{\rm m1} \le \theta_{\rm m1}$; 
that is, 
\begin{equation}
  n_{1}
  =
  2 W_{1} 
  {\rm e}^{\mathscr{F}(0,0)/(k_{\rm B}T)}
  \int_{0}^{\theta_{\rm m1}} {\rm d} \theta_{1}
  \sin\theta_{1}
  \exp
  \left[
    -\frac{\mathscr{F}(\theta_{1},0)}{k_{\rm B}T}
  \right]. 
\end{equation}
It should be noted that the exponential term 
in the integral rapidly decreases 
by changing $\theta_{1}$ from 0 to $\theta_{\rm m1}$. 
Then, we replace $\mathscr{F}(\theta_{1},0)$ 
by its Taylor series about $\theta_{1}=0$, 
keep the terms up to the second order of $\theta_{1}$, 
and replace the upper limit of the integral by $\infty$. 
The first term of Taylor series, 
$\partial \mathscr{F}(0,0)/\partial \theta_{1}$, 
is zero because $\theta_{1}=0$ corresponds to 
the local minimum of $\mathscr{F}$. 
$\sin\theta_{1}$ is approximated to $\theta_{1}$. 
Then, we arrive Eq. (\ref{eq:I_1}). 
The numer of the particle in region 2, 
$n_{2}$, is obtained in a similar way; 
that is, $n_{2}=2W_{2}{\rm e}^{\mathscr{F}(\pi,0)/(k_{\rm B}T)}I_{2}$, 
where $I_{2}$ is given by 
\begin{equation}
  I_{2}
  =
  {\rm e}^{-\mathscr{F}(\pi,0)/(k_{\rm B}T)}
  \frac{k_{\rm B}T}{\partial^{2}\mathscr{F}(\pi,0)/\partial\theta_{1}^{2}}.
\end{equation}

The particle flow from region 1 to region 2, $I_{1\to 2}$, satisfies
[see Eq. (\ref{eq:J_theta1})]
\begin{equation}
  \frac{\partial W}{\partial\theta_{1}}
  +
  \frac{1}{k_{\rm B}T}
  \frac{\partial \mathscr{F}}{\partial\theta_{1}}
  W
  =
  -\left(
    \frac{M_{1}V_{1}}{\alpha_{1}\gamma_{1}k_{\rm B}T}
  \right)
  \frac{I_{1\to 2}}{2\sin\theta_{1}}.
  \label{eq:I_m1_sub1}
\end{equation}
According to Brown \cite{brown63}, 
we assume that $I_{1\to 2}$ is independent of $\theta_{1}$. 
By multiplying ${\rm e}^{\mathscr{F}(\theta_{1},0)/(k_{\rm B}T)}$ to Eq. (\ref{eq:I_m1_sub1}) 
and integrating it over $[0,\pi]$, 
the left hand side of Eq. (\ref{eq:I_m1_sub1}) is reduced to 
\begin{equation}
\begin{split}
  &
  \int_{0}^{\pi} {\rm d}\theta_{1}
  \frac{\partial}{\partial\theta_{1}}
  W {\rm e}^{\mathscr{F}(\theta_{1},0)/(k_{\rm B}T)}
\\
  &=
  W(\pi,0) {\rm e}^{\mathscr{F}(\pi,0)/(k_{\rm B}T)}
  -
  W(0,0) {\rm e}^{\mathscr{F}(0,0)/(k_{\rm B}T)}
\\
  &=
  \frac{1}{2}
  \left(
    \frac{n_{2}}{I_{2}}
    -
    \frac{n_{1}}{I_{1}}
  \right),
\end{split}
\end{equation}
where we use the definitions of $I_{1}$ and $I_{2}$, 
i.e., $n_{1}=2W_{1}{\rm e}^{\mathscr{F}(0,0)/(k_{\rm B}T)}I_{1}$ 
and $n_{2}=2W_{2}{\rm e}^{\mathscr{F}(\pi,0)/(k_{\rm B}T)}I_{2}$. 
On the other hand, 
by using Taylor series of $\mathscr{F}(\theta_{1},0)$ about $\theta_{1}=\theta_{\rm m1}$, 
the right hand side of Eq. (\ref{eq:I_m1_sub1}) is approximated to 
$-[M_{1}V_{1}/(2\alpha_{1}\gamma_{1}k_{\rm B}T)]I_{1\to 2}I_{\rm m1}$, 
where $I_{\rm m1}$ is given by Eq. (\ref{eq:I_m1}). 
Thus, we obtain 
\begin{equation}
  \frac{1}{2}
  \left(
    \frac{n_{2}}{I_{2}}
    -
    \frac{n_{1}}{I_{1}}
  \right)
  =
  -\frac{M_{1}V_{1}}{2\alpha_{1}\gamma_{1}k_{\rm B}T}
  I_{1\to 2}
  I_{\rm m1}.
  \label{eq:I_m1_sub2}
\end{equation}
Similarly, 
the number of the particles in regions 2 and 3, $n_{2}$ and $n_{3}$, 
and the particle flow from the region 2 to region 3, $I_{2\to 3}$,
satisfy 
\begin{equation}
  \frac{1}{2}
  \left(
    \frac{n_{3}}{I_{3}^{\prime}}
    -
    \frac{n_{2}}{I_{2}^{\prime}}
  \right)
  =
  -\frac{M_{2}V_{2}}{2\alpha_{2}\gamma_{2}k_{\rm B}T}
  I_{2\to 3}
  I_{\rm m2}, 
  \label{eq:I_m1_sub3}
\end{equation}
where $I_{2}^{\prime}$, $I_{3}^{\prime}$, 
and $I_{\rm m2}$ are, respectively, given by 
\begin{equation}
  I_{2}^{\prime}
  =
  {\rm e}^{-\mathscr{F}(\pi,0)/(k_{\rm B}T)}
  \frac{k_{\rm B}T}{\partial^{2}\mathscr{F}(\pi,0)/\partial\theta_{2}^{2}},
\end{equation}
\begin{equation}
  I_{3}^{\prime}
  =
  {\rm e}^{-\mathscr{F}(\pi,\pi)/(k_{\rm B}T)}
  \frac{k_{\rm B}T}{\partial^{2}\mathscr{F}(\pi,\pi)/\partial\theta_{2}^{2}},
\end{equation}
\begin{equation}
  I_{\rm m2}
  =
  \sqrt{
    -\frac{2\pi k_{\rm B}T}{\partial^{2}\mathscr{F}(\pi,\theta_{\rm m2})/\partial\theta_{\rm 2}^{2}}
  }
  \frac{{\rm e}^{\mathscr{F}(\pi,\theta_{\rm m2})/(k_{\rm B}T)}}{\sin\theta_{\rm m2}},
\end{equation}
where $\theta_{\rm m2}=\cos^{-1}[-(H_{\rm appl}-H_{J2})/H_{\rm an2}]$.

By using Eqs. (\ref{eq:I_m1_sub2}), (\ref{eq:I_m1_sub3}) and 
the continuity equations of the particle flows, 
$\dot{n}_{1}=-I_{1\to 2}$, 
$\dot{n}_{2}=I_{1\to 2}-I_{2\to 3}$, 
and $\dot{n}_{3}=I_{2\to 3}$, 
we obtain Eq. (\ref{eq:transition_equation_weak_limit}). 
The switching probabilities per unit time, $\nu_{ij}$, are given by 
\begin{equation}
  \nu_{12}
  =
  \left(
    \frac{\alpha_{1}\gamma_{1}k_{\rm B}T}{M_{1}V_{1}}
  \right)
  \frac{1}{I_{1}I_{\rm m1}}, 
\end{equation}
\begin{equation}
  \nu_{21}
  =
  \left(
    \frac{\alpha_{1}\gamma_{1}k_{\rm B}T}{M_{1}V_{1}}
  \right)
  \frac{1}{I_{2}I_{\rm m1}},
\end{equation}
\begin{equation}
  \nu_{23}
  =
  \left(
    \frac{\alpha_{2}\gamma_{2}k_{\rm B}T}{M_{2}V_{2}}
  \right)
  \frac{1}{I_{2}^{\prime}I_{\rm m2}}, 
\end{equation}
\begin{equation}
  \nu_{32}
  =
  \left(
    \frac{\alpha_{2}\gamma_{2}k_{\rm B}T}{M_{2}V_{2}}
  \right)
  \frac{1}{I_{3}^{\prime}I_{\rm m2}}.
\end{equation}
The explicit forms of $\nu_{ij}$ are 
obtained by using $\mathscr{F}(\theta_{1},\theta_{2})$ 
and its derivative, 
and given by Eqs. (\ref{eq:f_12_weak})-(\ref{eq:delta_weak_sub}). 

By applying the negative current and magnetic field, 
which induce the magnetization reversal 
from $\mathbf{m}_{1},\mathbf{m}_{2}=+\mathbf{e}_{z}$ to 
$\mathbf{m}_{1},\mathbf{m}_{2}=-\mathbf{e}_{z}$, 
$\nu_{12}$ and $\nu_{23}$ become 
much larger than $\nu_{21}$ and $\nu_{32}$, respectively. 
Then, Eq. (\ref{eq:transition_equation_weak_limit}) 
can be approximated to 
$\dot{n}_{1}=-\nu_{12}n_{1}$, 
$\dot{n}_{2}=\nu_{12}n_{1}-\nu_{23}n_{2}$ 
and $\dot{n}_{3}=\nu_{23}n_{2}$. 
Then, the solutions of $n_{1}$, $n_{2}$, and $n_{3}$ 
with the initial conditions, 
$n_{1}(0)=1$, $n_{2}(0)=0$, and $n_{3}(0)=0$, 
are given by Eqs. (\ref{eq:n1_weak})-(\ref{eq:switching_rate_weak}).


\begin{thebibliography}{23}
\expandafter\ifx\csname natexlab\endcsname\relax\def\natexlab#1{#1}\fi
\expandafter\ifx\csname bibnamefont\endcsname\relax
  \def\bibnamefont#1{#1}\fi
\expandafter\ifx\csname bibfnamefont\endcsname\relax
  \def\bibfnamefont#1{#1}\fi
\expandafter\ifx\csname citenamefont\endcsname\relax
  \def\citenamefont#1{#1}\fi
\expandafter\ifx\csname url\endcsname\relax
  \def\url#1{\texttt{#1}}\fi
\expandafter\ifx\csname urlprefix\endcsname\relax\def\urlprefix{URL }\fi
\providecommand{\bibinfo}[2]{#2}
\providecommand{\eprint}[2][]{\url{#2}}

\bibitem[{\citenamefont{Slonczewski}(1996)}]{slonczewski96}
\bibinfo{author}{\bibfnamefont{J.~C.} \bibnamefont{Slonczewski}},
  \bibinfo{journal}{J. Magn. Magn. Mater.} \textbf{\bibinfo{volume}{159}},
  \bibinfo{pages}{L1} (\bibinfo{year}{1996}).

\bibitem[{\citenamefont{Slonczewski}(1989)}]{slonczewski89}
\bibinfo{author}{\bibfnamefont{J.~C.} \bibnamefont{Slonczewski}},
  \bibinfo{journal}{Phys. Rev. B} \textbf{\bibinfo{volume}{39}},
  \bibinfo{pages}{6995} (\bibinfo{year}{1989}).

\bibitem[{\citenamefont{Berger}(1996)}]{berger96}
\bibinfo{author}{\bibfnamefont{L.}~\bibnamefont{Berger}},
  \bibinfo{journal}{Phys. Rev. B} \textbf{\bibinfo{volume}{54}},
  \bibinfo{pages}{9353} (\bibinfo{year}{1996}).

\bibitem[{\citenamefont{Katine et~al.}(2000)\citenamefont{Katine, Albert,
  Buhrman, Myers, and Ralph}}]{katine00}
\bibinfo{author}{\bibfnamefont{J.~A.} \bibnamefont{Katine}},
  \bibinfo{author}{\bibfnamefont{F.~J.} \bibnamefont{Albert}},
  \bibinfo{author}{\bibfnamefont{R.~A.} \bibnamefont{Buhrman}},
  \bibinfo{author}{\bibfnamefont{E.~B.} \bibnamefont{Myers}}, \bibnamefont{and}
  \bibinfo{author}{\bibfnamefont{D.~C.} \bibnamefont{Ralph}},
  \bibinfo{journal}{Phys. Rev. Lett.} \textbf{\bibinfo{volume}{84}},
  \bibinfo{pages}{3149} (\bibinfo{year}{2000}).

\bibitem[{\citenamefont{Kiselev et~al.}(2003)\citenamefont{Kiselev, Sankey,
  Krivorotov, Emley, Schoelkopf, Buhrman, and Ralph}}]{kiselev03}
\bibinfo{author}{\bibfnamefont{S.~I.} \bibnamefont{Kiselev}},
  \bibinfo{author}{\bibfnamefont{J.~C.} \bibnamefont{Sankey}},
  \bibinfo{author}{\bibfnamefont{I.~N.} \bibnamefont{Krivorotov}},
  \bibinfo{author}{\bibfnamefont{N.~C.} \bibnamefont{Emley}},
  \bibinfo{author}{\bibfnamefont{R.~J.} \bibnamefont{Schoelkopf}},
  \bibinfo{author}{\bibfnamefont{R.~A.} \bibnamefont{Buhrman}},
  \bibnamefont{and} \bibinfo{author}{\bibfnamefont{D.~C.} \bibnamefont{Ralph}},
  \bibinfo{journal}{Nature} \textbf{\bibinfo{volume}{425}},
  \bibinfo{pages}{380} (\bibinfo{year}{2003}).

\bibitem[{\citenamefont{Huai et~al.}(2004)\citenamefont{Huai, Albert, Nguyen,
  Pakala, and Valet}}]{huai04}
\bibinfo{author}{\bibfnamefont{Y.}~\bibnamefont{Huai}},
  \bibinfo{author}{\bibfnamefont{F.}~\bibnamefont{Albert}},
  \bibinfo{author}{\bibfnamefont{P.}~\bibnamefont{Nguyen}},
  \bibinfo{author}{\bibfnamefont{M.}~\bibnamefont{Pakala}}, \bibnamefont{and}
  \bibinfo{author}{\bibfnamefont{T.}~\bibnamefont{Valet}},
  \bibinfo{journal}{Appl. Phys. Lett.} \textbf{\bibinfo{volume}{84}},
  \bibinfo{pages}{3118} (\bibinfo{year}{2004}).

\bibitem[{\citenamefont{Fuchs et~al.}(2004)\citenamefont{Fuchs, Emley,
  Krivorotov, Braganca, Ryan, Kiselev, Sankey, Ralph, Buhrman, and
  Katine}}]{fuchs04}
\bibinfo{author}{\bibfnamefont{G.~D.} \bibnamefont{Fuchs}},
  \bibinfo{author}{\bibfnamefont{N.~C.} \bibnamefont{Emley}},
  \bibinfo{author}{\bibfnamefont{I.~N.} \bibnamefont{Krivorotov}},
  \bibinfo{author}{\bibfnamefont{P.~M.} \bibnamefont{Braganca}},
  \bibinfo{author}{\bibfnamefont{E.~M.} \bibnamefont{Ryan}},
  \bibinfo{author}{\bibfnamefont{S.~I.} \bibnamefont{Kiselev}},
  \bibinfo{author}{\bibfnamefont{J.~C.} \bibnamefont{Sankey}},
  \bibinfo{author}{\bibfnamefont{D.~C.} \bibnamefont{Ralph}},
  \bibinfo{author}{\bibfnamefont{R.~A.} \bibnamefont{Buhrman}},
  \bibnamefont{and} \bibinfo{author}{\bibfnamefont{J.~A.}
  \bibnamefont{Katine}}, \bibinfo{journal}{Appl. Phys. Lett.}
  \textbf{\bibinfo{volume}{85}}, \bibinfo{pages}{1205} (\bibinfo{year}{2004}).

\bibitem[{\citenamefont{Hayakawa et~al.}(2008)\citenamefont{Hayakawa, Ikeda,
  Miura, Yamanouchi, Lee, Sasaki, Ichimura, Ito, Kawahara, Takemura,
  Meguro, Matsukura, Takahashi, Matsuoka, and Ohno}}]{hayakawa08}
\bibinfo{author}{\bibfnamefont{J.}~\bibnamefont{Hayakawa}},
  \bibinfo{author}{\bibfnamefont{S.}~\bibnamefont{Ikeda}},
  \bibinfo{author}{\bibfnamefont{K.}~\bibnamefont{Miura}},
  \bibinfo{author}{\bibfnamefont{M.}~\bibnamefont{Yamanouchi}},
  \bibinfo{author}{\bibfnamefont{Y.~M.} \bibnamefont{Lee}},
  \bibinfo{author}{\bibfnamefont{R.}~\bibnamefont{Sasaki}},
  \bibinfo{author}{\bibfnamefont{M.}~\bibnamefont{Ichimura}},
  \bibinfo{author}{\bibfnamefont{K.}~\bibnamefont{Ito}},
  \bibinfo{author}{\bibfnamefont{T.}~\bibnamefont{Kawahara}},
  \bibinfo{author}{\bibfnamefont{R.}~\bibnamefont{Takemura}},
  \bibinfo{author}{\bibfnamefont{T.}~\bibnamefont{Meguro}},
  \bibinfo{author}{\bibfnamefont{F.}~\bibnamefont{Matsukura}},
  \bibinfo{author}{\bibfnamefont{H.}~\bibnamefont{Takahashi}},
  \bibinfo{author}{\bibfnamefont{H.}~\bibnamefont{Matsuoka}}, \bibnamefont{and}
  \bibinfo{author}{\bibfnamefont{H.}~\bibnamefont{Ohno}},
  \bibinfo{journal}{IEEE. Trans. Magn.}
  \textbf{\bibinfo{volume}{44}}, \bibinfo{pages}{1962} (\bibinfo{year}{2008}).

\bibitem[{\citenamefont{Yakata et~al.}(2009)\citenamefont{Yakata, Kubota,
  Sugano, Seki, Yakushiji, Fukushima, Yuasa, and Ando}}]{yakata09}
\bibinfo{author}{\bibfnamefont{S.}~\bibnamefont{Yakata}},
  \bibinfo{author}{\bibfnamefont{H.}~\bibnamefont{Kubota}},
  \bibinfo{author}{\bibfnamefont{T.}~\bibnamefont{Sugano}},
  \bibinfo{author}{\bibfnamefont{T.}~\bibnamefont{Seki}},
  \bibinfo{author}{\bibfnamefont{K.}~\bibnamefont{Yakushiji}},
  \bibinfo{author}{\bibfnamefont{A.}~\bibnamefont{Fukushima}},
  \bibinfo{author}{\bibfnamefont{S.}~\bibnamefont{Yuasa}}, \bibnamefont{and}
  \bibinfo{author}{\bibfnamefont{K.}~\bibnamefont{Ando}},
  \bibinfo{journal}{Appl. Phys. Lett.} \textbf{\bibinfo{volume}{95}},
  \bibinfo{pages}{242504} (\bibinfo{year}{2009}).

\bibitem[{\citenamefont{Yakata et~al.}(2010)\citenamefont{Yakata, Kubota, Seki,
  Yakushiji, Fukushima, Yuasa, and Ando}}]{yakata10}
\bibinfo{author}{\bibfnamefont{S.}~\bibnamefont{Yakata}},
  \bibinfo{author}{\bibfnamefont{H.}~\bibnamefont{Kubota}},
  \bibinfo{author}{\bibfnamefont{T.}~\bibnamefont{Seki}},
  \bibinfo{author}{\bibfnamefont{K.}~\bibnamefont{Yakushiji}},
  \bibinfo{author}{\bibfnamefont{A.}~\bibnamefont{Fukushima}},
  \bibinfo{author}{\bibfnamefont{S.}~\bibnamefont{Yuasa}}, \bibnamefont{and}
  \bibinfo{author}{\bibfnamefont{K.}~\bibnamefont{Ando}},
  \bibinfo{journal}{IEEE. Trans. Magn.} \textbf{\bibinfo{volume}{46}},
  \bibinfo{pages}{2232} (\bibinfo{year}{2010}).

\bibitem[{\citenamefont{Koch et~al.}(2004)\citenamefont{Koch, Katine, and
  Sun}}]{koch04}
\bibinfo{author}{\bibfnamefont{R.~H.} \bibnamefont{Koch}},
  \bibinfo{author}{\bibfnamefont{J.~A.} \bibnamefont{Katine}},
  \bibnamefont{and} \bibinfo{author}{\bibfnamefont{J.~Z.} \bibnamefont{Sun}},
  \bibinfo{journal}{Phys. Rev. Lett.} \textbf{\bibinfo{volume}{92}},
  \bibinfo{pages}{088302} (\bibinfo{year}{2004}).

\bibitem[{\citenamefont{Li and Zhang}(2004)}]{li04}
\bibinfo{author}{\bibfnamefont{Z.}~\bibnamefont{Li}} \bibnamefont{and}
  \bibinfo{author}{\bibfnamefont{S.}~\bibnamefont{Zhang}},
  \bibinfo{journal}{Phys. Rev. B} \textbf{\bibinfo{volume}{69}},
  \bibinfo{pages}{134416} (\bibinfo{year}{2004}).

\bibitem[{\citenamefont{Apalkov and Visscher}(2005)}]{apalkov05}
\bibinfo{author}{\bibfnamefont{D.~M.} \bibnamefont{Apalkov}} \bibnamefont{and}
  \bibinfo{author}{\bibfnamefont{P.~B.} \bibnamefont{Visscher}},
  \bibinfo{journal}{Phys. Rev. B} \textbf{\bibinfo{volume}{72}},
  \bibinfo{pages}{180405} (\bibinfo{year}{2005}).

\bibitem[{\citenamefont{Jr}(1963)}]{brown63}
\bibinfo{author}{\bibfnamefont{W.~F.~B.} \bibnamefont{Jr}},
  \bibinfo{journal}{Phys. Rev.} \textbf{\bibinfo{volume}{130}},
  \bibinfo{pages}{1677} (\bibinfo{year}{1963}).

\bibitem[{com({\natexlab{a}})}]{comment1}
\bibinfo{note}{The spin transfer torque switching in the synthetic free layer
  is performed at room temperature using an Ru layer with a thickness of
  a few nm \cite{hayakawa08,yakata09,yakata10}. The spin diffusion length of Ru
  at 4.2 K is 14 nm \cite{bass07} and it would be much smaller at room
  temperature.}

\bibitem[{com({\natexlab{b}})}]{comment2}
\bibinfo{note}{Private communication with Hitoshi Kubota. It was experimentally
  shown that the critical current of the spin transfer torque switching in a
  CoFeB/Ru/CoFeB spin-valve is one order of magnitude larger than that in
  CoFeB/MgO/CoFeB MTJs (unpublished). This result means that the spin transfer
  torque arising between the free layers of the synthetic structure is
  negligible compared to that arising from the spin current injected from the
  fixed layer.}

\bibitem[{\citenamefont{Brataas et~al.}(2001)\citenamefont{Brataas, Nazarov,
  and Bauer}}]{brataas01}
\bibinfo{author}{\bibfnamefont{A.}~\bibnamefont{Brataas}},
  \bibinfo{author}{\bibfnamefont{Y.~V.} \bibnamefont{Nazarov}},
  \bibnamefont{and} \bibinfo{author}{\bibfnamefont{G.~E.~W.}
  \bibnamefont{Bauer}}, \bibinfo{journal}{Eur. Phys. J. B}
  \textbf{\bibinfo{volume}{22}}, \bibinfo{pages}{99} (\bibinfo{year}{2001}).

\bibitem[{\citenamefont{Zhang et~al.}(2002)\citenamefont{Zhang, Levy, and
  Fert}}]{zhang02}
\bibinfo{author}{\bibfnamefont{S.}~\bibnamefont{Zhang}},
  \bibinfo{author}{\bibfnamefont{P.~M.} \bibnamefont{Levy}}, \bibnamefont{and}
  \bibinfo{author}{\bibfnamefont{A.}~\bibnamefont{Fert}},
  \bibinfo{journal}{Phys. Rev. Lett.} \textbf{\bibinfo{volume}{88}},
  \bibinfo{pages}{236601} (\bibinfo{year}{2002}).

\bibitem[{\citenamefont{Oogane et~al.}(2006)\citenamefont{Oogane, Wakitani, 
  Yakata, Yilgin, Ando, Sakuma, and Miyazaki}}]{oogane06}
\bibinfo{author}{\bibfnamefont{M.}~\bibnamefont{Oogane}},
  \bibinfo{author}{\bibfnamefont{T.} \bibnamefont{Wakitani}},
  \bibinfo{author}{\bibfnamefont{S.} \bibnamefont{Yakata}},
  \bibinfo{author}{\bibfnamefont{R.} \bibnamefont{Yilgin}},
  \bibinfo{author}{\bibfnamefont{Y.} \bibnamefont{Ando}},
  \bibinfo{author}{\bibfnamefont{A.} \bibnamefont{Sakuma}}, \bibnamefont{and}
  \bibinfo{author}{\bibfnamefont{T.}~\bibnamefont{Miyazaki}},
  \bibinfo{journal}{Jpn. J. Appl. Phys.} \textbf{\bibinfo{volume}{45}},
  \bibinfo{pages}{3889} (\bibinfo{year}{2006}).

\bibitem[{\citenamefont{Suzki et~al.}(2009)\citenamefont{Suzki, Tulapurkar, and
  Chappert}}]{suzuki09}
\bibinfo{author}{\bibfnamefont{Y.}~\bibnamefont{Suzki}},
  \bibinfo{author}{\bibfnamefont{A.~A.} \bibnamefont{Tulapurkar}},
  \bibnamefont{and} \bibinfo{author}{\bibfnamefont{C.}~\bibnamefont{Chappert}},
  \emph{\bibinfo{title}{Nanomagnetism and Spintronics}}
  (\bibinfo{publisher}{Elsevier}, \bibinfo{year}{2009}),
  \bibinfo{note}{{C}hapter 3}.

\bibitem[{com({\natexlab{c}})}]{comment3}
\bibinfo{note}{Private communication with Hitoshi Kubota and Shinji Yuasa.}

\bibitem[{\citenamefont{Bass and W.~P.~Pratt}(2007)}]{bass07}
\bibinfo{author}{\bibfnamefont{J.}~\bibnamefont{Bass}} \bibnamefont{and}
  \bibinfo{author}{\bibfnamefont{J.}~\bibnamefont{W.~P.~Pratt}},
  \bibinfo{journal}{J. Phys.: Condens. Matter} \textbf{\bibinfo{volume}{19}},
  \bibinfo{pages}{183201} (\bibinfo{year}{2007}).

\bibitem[{com({\natexlab{c}})}]{comment4}
\bibinfo{note}{Private communication with C. Mewes. 
               Their results were recently presented at 
               55${}^{\rm th}$ Annual Conference on Magnetism and Magnetic Materials 
               by W. Butler (HC-09). }

\end{thebibliography}
\end{document}